\documentclass[10pt, conference, letterpaper]{IEEEtran}

\usepackage{graphicx}
\usepackage{graphics}
\usepackage{subfigure,subfigmat}

\def\BibTeX{{\rm B\kern-.05em{\sc i\kern-.025em b}\kern-.08emT\kern-.1667em\lower.7ex\hbox{E}\kern-.125emX}}
    
\begin{document}

%
\title{Consumer UAV Cybersecurity Vulnerability Assessment Using Fuzzing Tests}

%


%

\author{
\IEEEauthorblockN{
    David Rudo\IEEEauthorrefmark{1} and
Kai~Zeng\IEEEauthorrefmark{2}}
\IEEEauthorblockA{
    \IEEEauthorrefmark{1}Centreville High School, 5609 Willow Valley Rd, Clifton, VA 20124\\
    \IEEEauthorrefmark{2}Wireless Innovation and Cybersecurity Lab, ECE Department\\ George Mason University, Fairfax, VA 22030\\
    Email:~davidrudo@hotmail.com, kzeng2@gmu.edu
   }
}

\maketitle

%
\begin{abstract}
 Unmanned Aerial Vehicles (UAVs) are remote-controlled vehicles capable of flight and are present in a variety of environments from military operations to domestic enjoyment. These vehicles are great assets, but just as their pilot can control them remotely, cyberattacks can be executed in a similar manner. Cyber attacks on UAVs can bring a plethora of issues to physical and virtual systems. Such malfunctions are capable of giving an attacker the ability to steal data, incapacitate the UAV, or hijack the UAV.  To mitigate such attacks, it is necessary to identify and patch vulnerabilities that may be maliciously exploited. In this paper, a new UAV vulnerability is explored with related UAV security practices identified for possible exploitation using large streams of data sent at specific ports. The more in-depth model involves strings of data involving FTP-specific keywords sent to the UAV's FTP port in the form of a fuzzing test and launching thousands of packets at other ports on the UAV as well. During these tests, virtual and physical systems are monitored extensively to identify specific patterns and vulnerabilities. This model is applied to a Parrot Bebop 2, which accurately portrays a UAV that had their network compromised by an attacker and portrays many lower-end UAV models for domestic use. During testings, the Parrot Bebop 2 is monitored for degradation in GPS performance, video speed, the UAV's reactivity to the pilot, motor function, and the accuracy of the UAV's sensor data. All these points of monitoring give a comprehensive view of the UAV's reaction to each individual test. In this paper, countermeasures to combat the exploitation of this vulnerability will be discussed as well as possible attacks that can branch from the fuzzing tests.
\end{abstract}

\thispagestyle{empty}
%

\section{Introduction}
In recent years, unmanned aerial vehicles (UAVs) have taken their foothold in our society whether it is for domestic or military use. UAVs are IoT devices that users can pilot for a multitude of purposes. UAVs have been gaining popularity specifically in households, where they are becoming more and more commonplace. While UAVs that are used by professional cinematographers or military personnel may have extremely robust security systems, many UAVs made for consumer purchase are still extremely vulnerable. Many consumer UAVs have similar vulnerable features, no WPA2 encryption for network security by default, and a free or open-source mobile application to pilot the UAV. These two features create a targetable access point for malicious attacks to be launched upon. Due to such limited cybersecurity put into many consumer UAVs, there has already been plenty of research done on the multiple types of exploits that can be executed for malicious intent. There are three main types of attacks that an attacker can use on a UAV: data theft, hijacking, and incapacitation, which are all very dangerous types of attacks that specifically consumer drones commonly fall victim to.  \cite{krishna, Xingqin, usenix, networks}.

The main attribute all of these types of attacks have is their remote execution. Many drone attacks such as replay, injection, hijacking, and delay all involve weakening a specific subsystem. UAVs must manage much more information than just the user's input to be a competitive product in the market. The special features included in many consumer UAVs lead to more ways that products can be exploited. Many UAV attacks are based around weakening the GPS system due to it being on almost all major UAV products. GPS based attacks also allow the user to control what the UAV sees and interacts with, giving the attacker direct access to how the UAV communicates with the world around it. The fuzzing test talked about in this paper is a GPS based attack as well, and may be expanded and implemented for the same purposes.

\begin{figure}[t!]
  \includegraphics[width=0.625\linewidth]{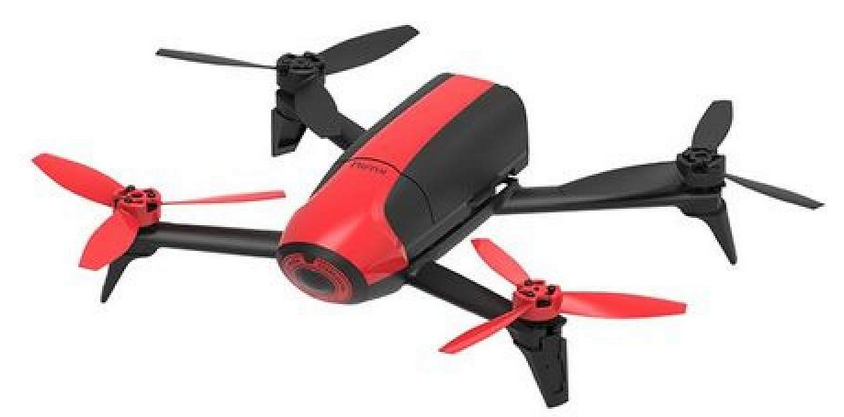}
  \caption{Parrot Bebop 2 drone used for testing}
  \label{fig:Parrot Bebop 2}
\end{figure}

The fuzzing test discussed in this paper was launched on a Parrot Bebop 2, shown in \ref{fig:Parrot Bebop 2}, a drone equipped with autonomous path settings, GPS navigation, an altimeter, an accelerometer, and a gyroscope. With a mobile app to control the drone and no WPA2 encryption on the network, the Parrot Bebop 2 is a good standard for a household drone. The fuzzing test involves the exploitation of FTP anonymous and other weak ports when subjected to fuzzing and flooding attacks. The model discussed in this paper can also be used to diagnose other UAVs on the market to see if they fall victim to this attack in a similar, if not exact, fashion. The attack does not only expose a way to jam communication with the UAV's GPS, but demonstrates new methodologies to launch common UAV attacks and exploit other IoT devices.

This new vulnerability, discovered through fuzzing tests, is a clear example of what other methods of cybersecurity need to implemented in consumer UAVs and other IoT devices. Maintaining port security and implementing WiFi protocols such as WPA2, are important measures that can be taken to combat the exploitation of this vulnerability. For IoT devices and UAVs these countermeasures can be implemented simply and effectively fend off attacks. Cybersecurity related countermeasures that involve how data is managed and processed is also important when patching this vulnerability. Using basic computer science strategies to immediately detect malicious fuzzing tests. The novel effect of GPS communication being forcefully terminated is a major security risk and is why such countermeasures should be used. 

\section{Related Work}
\subsection{Overview of UAV Security}
As more devices become integrated with local networks, the importance of understanding the proper and efficient ways to secure internet of things (IoT) devices greatens. The basic operation of UAVs, malicious physical attacks on UAVs, malicious logical attacks on UAVs, UAV forensics, and secure communication with UAVs are topics of critical importance to UAV security. Many UAVs are developed to function autonomously and in large networks. With this technology prevalent, advancements in information security need to be made for the safe operation of large, autonomous UAV networks. Current developments in information security for UAVs involve encryption methods and new network models, but that barely considers external threats to security.

Before discussing the cybersecurity aspects of UAVs, the way UAVs operate must be understood. UAVs use GSM (Global System for Mobile Communication) to maintain network connectivity. UAVs also use UTM platforms, UTM platforms are unmanned traffic management platforms that are used to secure low-level operations. UTM platforms work by pinging the UAV specific commands based on the UAV's proximity to the platform. The UTM platform is used for navigating autonomously, landing, and indicating locations for specific tasks. Within this network, UAVs must overcome the challenges of collision, privacy invasion, and information security intrusion. Seven key components must be exceptional for UAVs to avoid failing these challenges: Secure pilot registration, UTM data security, UAV data security, secure data exchange, secure command transmittance, reliable tracking, and secure network connectivity. These components are critical to UAV functionality. 

\subsection{Physical Attacks}
With an understanding of how a UAV operates, it is worth discussing how practical operations can be abused, interrupted, and intercepted by external threats  \cite{infosec} \cite{twente}. While many people fear the idea of logical attacks on UAVs, physical attacks are also a possibility. Three types of physical attacks that are used on UAVs are projectiles, interception drones, and microwave beam jamming. The projectile method involves expertise sharpshooting or use of missiles; it is not an ideal method for stealth or when in a public area. The interception method involves a drone releasing a net or projecting wire to incapacitate another drone in the area. The final method is only used by the military and involves the projection of microwave beams at a UAV to disrupt communication signals. While these physical attacks are extremely effective their counterparts are much more common

\subsection{Logical Attacks}
Logical attacks are a huge concern with UAVs due to their prevalence in the IoT \cite{infosec, twente, Hartmann}. When an IoT device becomes very common, it leads to shoddier brands with many vulnerabilities. Many attackers rely on this concept to exploit UAV security and gather private user information. One exploit that takes advantage of less expensive UAVs is when a hacker uses forced signals to target the GPS. This attack only works with low-cost UAVs that have no application layer encryptions and no authentication mechanisms. When successful, this gives the attacker complete control of the UAV's actions. Another exploit that only works on commercial UAVs is being able to hijack video feed by collecting the video fragments on low-frequency channels and then reconstructing those fragments into a full video. More expensive UAVs don't fragment their video into multiple lower frequency channels.  Two attacks that do affect more expensive UAVs are DoS attacks and replay attacks. A DoS attack involves flooding all channels of the UAV, restricting the signal between the pilot and the UAV. DoS attacks can also be conducted by sending deauthentication packets on the same Wi-Fi channel. A replay exploit involves recording commands intercepted from the pilot and UAV's communication channel. After the signals have been recorded, the attacker replays the recording to the UAV to test packets of data. UAV vulnerabilities can also be exploited by other UAVs rather than a computer. An organization called Hak5 was able to use a UAV that could force Parrot AR Drones to fly in failsafe mode with the tools Pineapple Wi-Fi and BatterPack installed. Hak5 also developed a UAV that functioned as a flying Wi-Fi sniffer. 

\subsection{Forensics}
With every vulnerability exploited there are cyberforensics analysts that must understand the extent of the attack. Cyberforensics with UAVs is no exception, there are many things an analyst needs to find to understand the source of the issue. If an analyst is retrieving data after an incident, it is typical to take a forensic image of the UAV's SSD. With this forensic image, an investigator can find what commands were given, what the UAV was streaming, the GPS location, and the network communication that took place. This data is stored in logs with a .dat or .txt file extension. Forensic investigators can also use real-time tracking to monitor the activity taking place. When a cyberforensics specialist is tracking a UAV in real-time, sensor data must be closely monitored. Real-time tracking is done by connecting to the UAV via Wi-Fi and sending commands for data dumps and status updates. If monitored with SSH, a forensic analyzer can get the UAV to dump root files through the connection. Forensics analyzers also use Telnet to monitor UAV connections due to Telnet's ability to send an image of the root directory. \cite{sans, drop}

\subsection{Fuzzing}
A UAV's network security has also been profoundly researched. Many UAVs have unrestricted communication systems. Services such as FTP, Telnet, and SSH have all been used to steal information. Specifically, FTP anonymous has been exploited on UAVs using FileZilla. Without any proper authentication, an attacker is able to download the entire filesystem of the UAV's FTP server which contains flight logs, video, GPS data, and sensor readings. Such unsecured communication systems have been victims of numerous exploits. Fuzzing, floods, and DoS attacks have all been used to exploit the weaknesses in unsecured communication systems. Floods and DoS attacks have been used to crash UAVs and steal private information from users, while fuzzing has been used to cause erratic behavior in UAVs. 

\subsection{IoT Networks}
During the entire process of monitoring, exploiting, or protecting UAVs, a certain level of network security is involved. A secure channel for communication is critical for all IoT devices. The current structure for many IoT networks is simplistic and insecure. The current IoT network used is a network connecting a physical machine with a digital medium to create data. In the instance of a drone, the drone itself is the hardware, the embedded Linux or OS is the digital medium, and the maintenance log is the data created. Between these exchanges of information, there is very little security and plenty opportunity for interception and interruption without proper management and encryption. 

\section{Testing Methodology And Results}
\subsection{Hardware and Tools}
The new vulnerability discovered involves a fuzzing test on port 21 of the Parrot Bebop 2. The hardware used for this testing involved a Parrot Bebop 2 drone and a laptop with Kali Linux (2019) installed. The tool used to exploit the UAV's navigation system was the Metasploit ftp\_pre\_post auxiliary module located under the "/fuzzing/ftp" directories. The only alterations made to the code of this module involved isolating one specific command or a group of special characters. A fuzzing test also mentioned in this paper is conducted through a tool called InviteFlood, which is installed on Kali Linux by default. To run InviteFlood over a series of ports autonomously, a bash script, as seen in figure 5, is used to facilitate testing. For more information on the InviteFlood visit the manual page for in-depth detail about the arguments it needs. 

\subsection{Fuzzing Tests and FTP Anonymous}
A fuzzing test is a type of test where a stream of packets is sent at a machine while the size of the packet is incremented with the succession of each test \cite{hackBebop,traffic}. 
This novel test uses Metasploit's ftp\_pre\_post module to execute such a test. A very well known vulnerability of domestic UAVs is that they allow anonymous FTP. What this means is that anyone is allowed to log in and begin accessing the UAV's data. Media files, flight logs, and internal measurements can all be viewed without ever entering any credentials. While data is public with anonymous FTP, most UAV's don't allow edits or permission changes to the data. Although such security restrictions are typically put in place, the UAV still needs to process the commands directed toward the FTP server.

\subsection{FTP Commands}    
The FTP documentation gives further information on how each FTP command reacts with the UAV and its purpose. Some commands are straight-forward, like PASS, USER, and ACCT, that give user credentials or attempt logging in. Others are much more complicated like CWD, CDUP, and SMNT that control actions inside the file system. The documentation identifies four different types of commands that are viable for this test: User actions, file actions, requests, and memory. User actions are commands involving the users logging in or controlling functions from a high level. Commands like USER, PASS, ACCT, QUIT, REST, REIN, and ABORT all deal with either a user logging in, canceling, or restarting a specific action involving data management. File actions are commands that deal with the directories within the server. CWD, CDUP, SMNT, DELE, RMD, MKD, PWD, NLST, and RNTO either access, view, rename or delete files and folders within an FTP server. Request commands are commands that ask the FTP server for a specific piece of data. The commands PORT, PASV, TYPE, STRU, RETR, LIST, NLST, NOOP, SITE, STATUS, and SYST all require the server to respond with either a message or send specific data about the filesystem or server. The final type of command is memory commands. Memory commands allocate space inside the server and request the server to store items and append stored data with new data. The commands STOU, APPE, STOR, ALLO, and RETR all require memory allocation with certain items to be replaced or combined.

\begin{figure}[!t]
    \includegraphics[width=\linewidth]{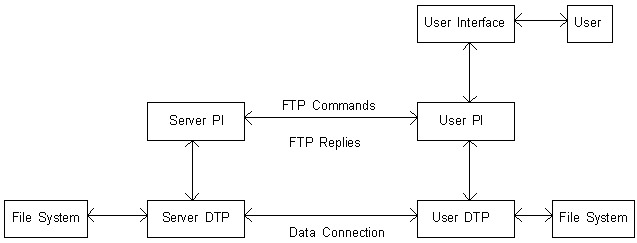}
    \caption{The basic structure of an FTP network connection \cite{ftp}}
    \label{fig:FTP network diagram} 
\end{figure}

\subsection{Metasploit FTP Fuzzer Module}
\begin{figure}[t!]
  \includegraphics[width=\linewidth]{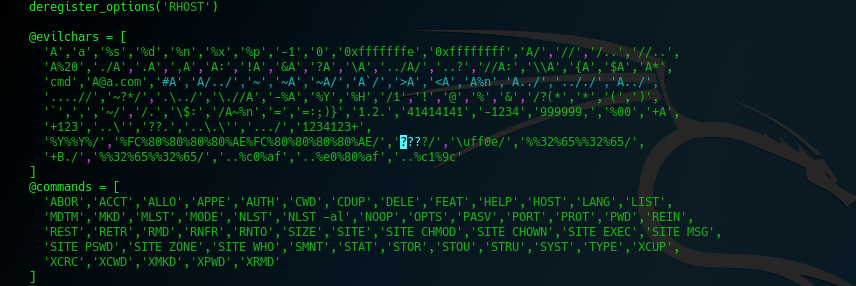}
  \caption{The two arrays in the Metasploit ftp\_pre\_post module with one containing FTP commands while the other stores special characters}
  \label{fig:Metasploit script arrays}
\end{figure}

To understand how these tests are launched, the ftp\_pre\_post module must be understood. The ftp\_pre\_post module traditionally works by sending packets in cycles. The Ruby script connects to the anonymous FTP service, provided by the Parrot, and becomes part of a network shown in Fig. \ref{fig:FTP network diagram}. It goes through a list of commands shown in Fig. \ref{fig:Metasploit script arrays} and then sends those commands with an increasing packet size in bytes. It loops through these commands until it reaches the last one. It will then repeat the same process but with special characters also included in the packet. The problem with this process is that it is hard to distinguish if a certain combination of commands, sizes, and special characters is more effective than another. To solve this issue the ruby script making up the module was edited in a way to isolate all elements. 

\begin{figure}[t!]
    \includegraphics[width=\linewidth]{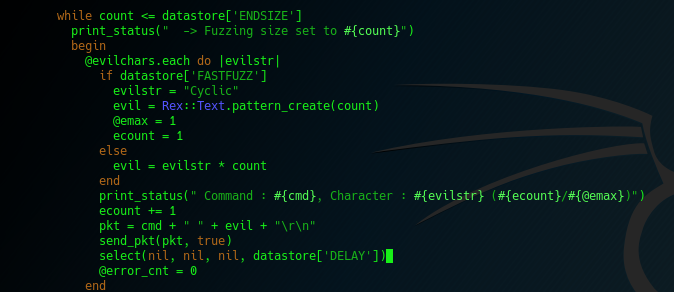}
    \caption{The enhanced for-loop that is running every combination of commands and special characters in the Metasploit ftp\_pre\_post module}
    \label{fig:Metasploit for-loop}
\end{figure}

The edited code, on a computer science level, is extremely basic. Essentially, it is two arrays of strings being looped through. One array contains multiple FTP commands and the other has special characters. These arrays are then looped through, in the form of nested for-loops shown in Fig. \ref{fig:Metasploit for-loop}, to create the maximum number of combinations. The code will then send the packet with the specified combination and will wait until the device responds back. Any major response is reported back to the user on the terminal. The edits to the code mainly involve those two arrays. Deleting and adding certain keywords and special characters to look at the individual effects of each characteristic. With such edits in place, testing the specifics of each command and special character became a much easier process.

\subsection{Testing Overview}
\begin{figure}[t!]
  \centering
  \subfigure[Normal display when piloting the Parrot Bebop 2]{
    \includegraphics[width=0.4\textwidth]{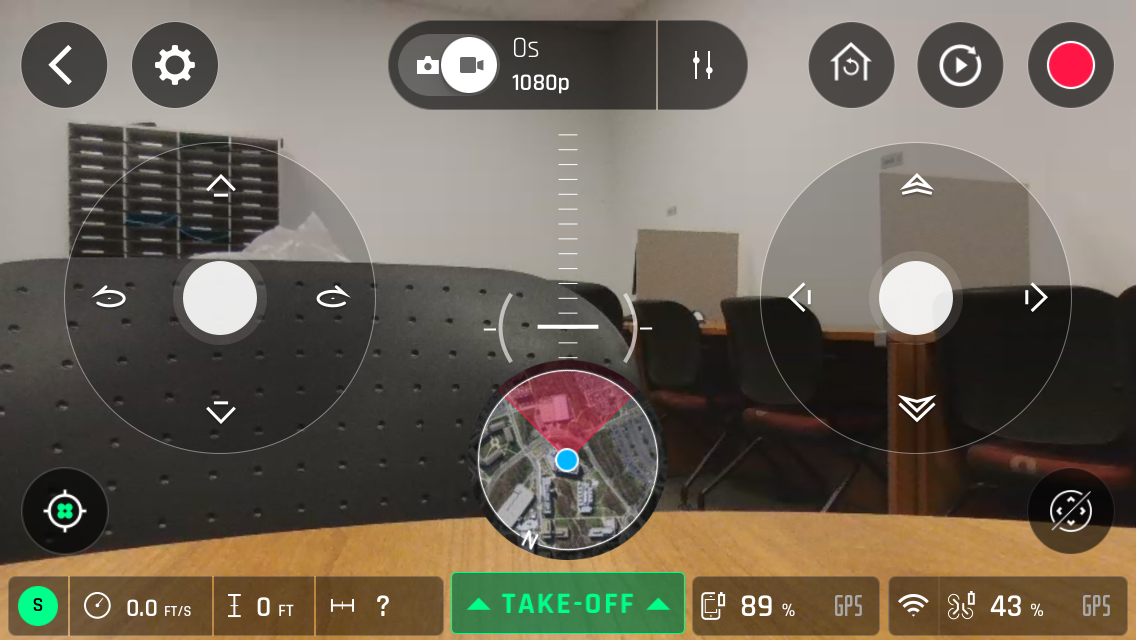}
    }
  
  \subfigure[Pilot view on mobile app when UAV is disconnected]{
    \includegraphics[width=0.4\textwidth]{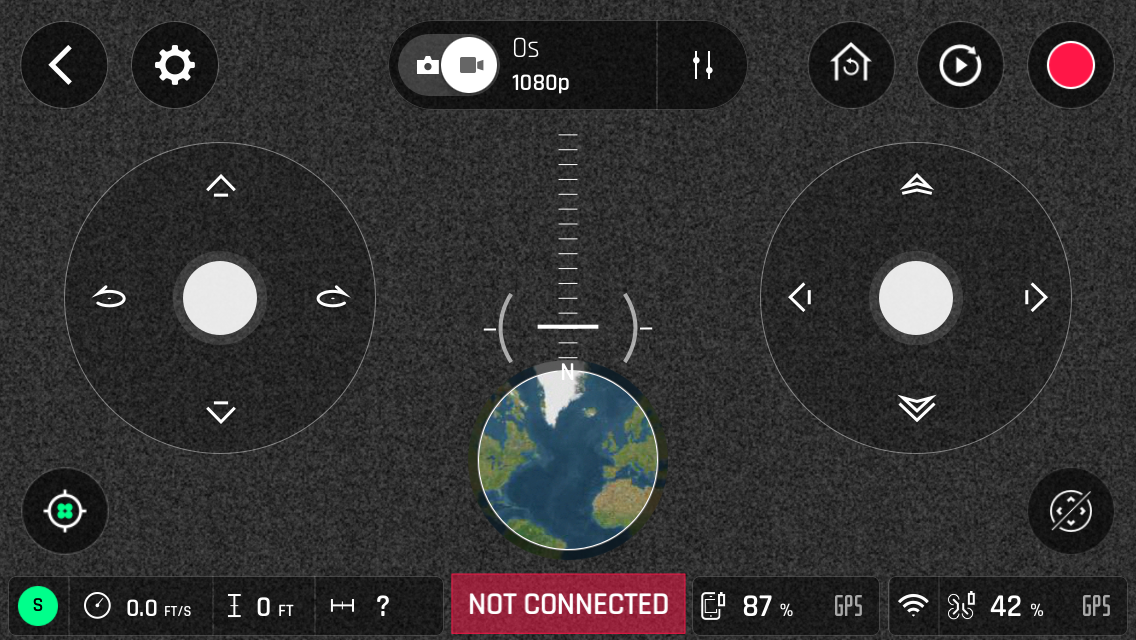}
    }
    \caption{The blue globe is an indicator that the GPS function is unavailable. When a fuzzing test occurs, the video feed in [a] will be displayed with [b]'s blue globe instead}
  \label{fig:Parrot_Interface}
\end{figure}

To see if two different types of commands made a difference when fuzzing, two very different commands were tested ABOR and REIN. ABOR and REIN's differing functions was a good way to test if a wide variety of commands would have an effect during a fuzzing test. 
The results of each test are monitored through the Parrot's mobile application shown in Fig. \ref{fig:Parrot_Interface}. Any hints the mobile application gave of critical services disconnecting were monitored carefully and recorded accordingly.

\begin{table}[t!]
\begin{center}
 \begin{tabular}{||p{0.5in} p{1in} p{0.7in} p{0.5in}||} 
 \hline
 FTP Command & Description of Command & Basic Range (Bytes) & Range w/ Special Characters (Bytes) \\ [0.5ex]
 \hline\hline
 ABOR & Aborts a task or operation &100-10,000 & -\\ 
 \hline
 PASS & Specify the user's password for logging in &100-10,000 & -\\
 \hline
 ACCT & Lists user account information &100-10,000 & -\\
 \hline
 QUIT & Closes current connection &100-10,000 & -\\
 \hline
 REST & Tells server to prepare for restart&100-10,000 & -\\  
 \hline
 REIN & Re-initializes current connection &7.500-8,500 & -\\ 
 \hline
 PORT & Lists active ports being used &100-1,000 & 100-5,000\\
 \hline
 PASV & Begins a service on port waiting to be started by the user &100-1,000 & 100-10,000\\
 \hline
 RETR & Retrieve a file or value &2,000-3,000 & 100-10,000\\
 \hline
 CWD & Change the directory &1-3,000 & 3,300-3,750\\
 \hline
 CDUP & Go to previous directory &2,800 & 100-1,300\\
 \hline
 SMNT & Mount a different file system structure on the UAV &1,000-5,000 & -\\
 \hline
 DELE & Delete a file or folder &100-10,000 & -\\[1ex]
 \hline
\end{tabular}
\end{center}
\caption{FTP commands used in fuzzing tests}
\label{t:FTP_commands}
\end{table}

ABOR, which is short for abort, is a command that tells the server to abort a previous FTP command or process. All data transfer is forced to be shut down and all live connections involving the transfer of data must be terminated. REIN is much different, involving the reinitialization of the current user by terminating the current login. REIN flushes all account information except to allow any transfer of data still in progress. This command is typically used immediately before logging in. Both commands, being very different, showed very different results. ABOR showed major video obscuration with packet sizes in the range of 3,000 to 10,000 bytes. The video obscuration is most noticeable when the drone is moving at high speeds and the level of obscuration would spike with special characters like "\%x". After reaching a packet size larger than 10,000 bytes no more effects are shown on the Parrot. REIN showed little to no video obscuration in comparison to ABOR. Instead, it would block the drone's GPS from displaying its location. With packet sizes ranging from 7,500 to 8,500 bytes, the drone would be unable to locate or self-navigate. The REIN command also drains the battery of the drone dramatically. Special characters were ineffective since REIN is a command that never takes any parameters. Both commands did have one outcome in common, the motors of the UAV losing control. This symptom, the video obscuration, and the battery losing power rapidly all appear when the test is launched during flight. As the UAV loses control of its motors and begins swirling downwards, the video feed will break into static or be extremely delayed, and the UAV's battery, in some cases, will drain in a matter of minutes. Such symptoms would incapacitate a user's ability to pilot the UAV effectively and is a clear threat. All of these effects, while effective in taking down the UAV, are results of an overflow attack. The UAV is unable to handle such large streams of data so quickly and begins draining power from other resources and puts less effort in other processes. This explains the drain in battery power, the video feed cutting, and the functionality of the motors declining.

The more interesting, and novel, effect is the GPS cutting out. With specific commands and special characters, it was possible to consistently have the Parrot lose its GPS signal. This typically involves fuzzing the drone with an FTP command that deals with file-based operations such as changing directories or retrieving data. The first thing known about these incidences is that they are not overflow attacks in terms of the size of the packets. This is proven by a test resulting in the GPS to cut out on a packet with a size of 100 bytes when the command CWD or CDUP was used with the "/" special character. The packet size is too small for the Parrot to be overloaded. As for when there are large streams of data, the tests become slightly harder to monitor. The size may not be at a level for an overflow attack to occur, but the UAV will clearly process and accept packets that don't fit within the parameters of the command. This prolongs the lack of GPS functionality extensively and can lead to certain commands going past the UAV, or crashing the drone entirely. Certain commands have also been tested with specific triggering parameters used. In such a test case there were only a couple packets that didn't fit the normal parameters accepted. Not only were the incorrect parameters overrun easily, but the range at which the GPS would cut out varied differently than when the command was sent without any special characters.  

An important fact to know about this GPS cut out test is that each command behaves differently. Each command used in the test will vary in terms of what size the packets must be to be effective, the best complementing special characters, the longevity of the effects, the consistency of the effects, and any extraneous effects the UAV displays. Many of these differences have to deal with security features hard-coded into the Parrot. For instance, if a developer knows the range of sizes for a proper packet of data regarding a command, then the developer can set a minimum and maximum limit for the size of the packet to be before processing. This means if a packet is sent and is above or below the limit, the UAV will automatically reject that piece of data and won't bother processing it. The ranges will also change depending on if there is a parameter, or special character in this case, that is called in conjunction with the command. These limits don't just affect how large a packet of data can be. As mentioned before, the autonomous FTP service on most UAVs doesn't allow the filesystem to be edited. This means that as soon as a restricted process is detected by the server, the request will immediately be rejected. What testing has shown, however, is that the FTP server will process what the command is requesting in its entirety before deciding if such an action is permissible and should be rejected.

\subsection{User Action Commands}
The first group of commands worked on when studying this novel GPS test, were the user actions. With exception of one command, most user actions had limited to no effects on the UAV's functionality. Commands that involved logging in the user or terminating processes such as USER, PASS, ACCT, QUIT, and REST would either terminate the connection with the FTP server or have negligible effects. The only exception to this group of commands when testing was REIN. REIN's success compared to the other commands in its family is credited to the fact that REIN doesn't involve logging in the user nor does it disrupt any current processes running on the UAV. What makes REIN a very interesting command, is that it can stack the same request over and over for the UAV to execute without affecting the current request being handled. REIN, although showing quite minimal video obscuration, was capable of disrupting the GPS signal when packet size was in the range of 7,500 to 8,500 bytes. REIN was also able to quickly drain the battery of the UAV in a matter of 5 minutes when all the packets sent followed the UAV's configured packet size range. REIN isn't used with any set of parameters, which is why special characters made no difference while testing.

\subsection{Request Memory Commands}
The next tested group of commands was the request and memory commands due to their similarity. The memory command RETR and the request commands PORT and PASV were three diverse commands that could fairly represent the less common commands used. The memory command, RETR is an abbreviation of the word retrieve. The RETR command takes a parameter of a pathname and then retrieves the data located in that path name for the user. The RETR command can also be used in conjunction with the other memory commands and the REST command. The command's main purpose is to transfer a copy of a specified file while leaving the status of the file unchanged. PORT is a command that retrieves what port should be used in a data connection. The command will list the defaults for both the user and server in the form of a concatenated 32-bit internet host address and 16-bit TCP port address. PORT will also accept parameters with the intention to give more information on a specific user, server, or port number. The PASV command, which differs greatly from the PORT command, requests the server to listen on data port. The port the server listens to is not the default data port, and the server must wait on that port until a connection is requested by a user. The PASV command also outputs the host and port address being listened on. Similar to the PORT and RETR command, the PASV command also takes a parameter and in this case, it is used to specify what port to listen on.

\subsubsection{RETR Command}
The RETR command was the first command tested to be more effective when parameters were used. With a packet size range of 2,000 to 3,000 bytes, the RETR command can cause GPS failure on its own. When used with non-alphabetic parameters, such as \%x, \%s, \%d, or \%n, the GPS function could crash at almost any range, giving the RETR command a much wider range for testing. These results most likely have to do with how the FTP command is storing these special characters. When it tries processing what the command is specifically looking for it is unable to handle the unknown character and begins pulling power away from other subsystems to process this request. When these similar RETR commands are sent to the UAV constantly, the GPS subsystem loses all power entirely. When looping through the entire array of special characters, the RETR command shuts down the GPS if special characters such as \%x, \%s, or \$\^, are used as parameters. However, the effects would immediately dissipate when alphabetic or file system characters were utilized. The reason for a dramatic change in the UAV's behavior is because alphabetic and file system characters are normal parameters for the RETR and are easy to process relative to the other special characters.

\subsubsection{PASV Command}
The tests with the PASV command shows how requests that require large amounts of computation fair in fuzzing tests. PASV, by itself, can take down the Parrot's GPS functionality when the size is in the range from 100 to 1,000 bytes. When parsing through the array of special characters, the PASV test began shutting down the Parrot's GPS communication after the set of "\%"-related characters. Other special characters had the same results, and the test continued until a much later special character in the list was used. Due to there being no uniqueness with the special character causing the test to end, this test's longevity has to do with how the UAV manages the PASV request. The documentation of the PASV command specifies that the server must wait for a connection on a specific server. Since this command is allowed to be called by the user and takes a numerical parameter, the Parrot will process and run the PASV command every time. With so many requests piling on top of each other, the Parrot is forced to put all of its resources towards the incoming flow of packets. This leads to a delay which allows valid requests to be ignored or skipped while the Parrot is trying to catch up. Such a test has a possibility to allow a more malicious command to be processed by the Parrot without proper authentication.

\subsubsection{PORT Command}
The PORT command, similar to RETR, would shut down the Parrot's GPS when the special characters \%s, \%d, \%n, \%x, or \%p were inputted. Unlike the RETR, the PORT command only affected the Parrot's GPS when special characters starting with "\%" were used. The GPS was only incapacitated for a relatively short amount of time and had minimal effects on the Parrot's general behavior. Compared to the processes demanded by PASV and RETR, PORT doesn't require much computational work for the UAV and explains why it isn't as effective as PASV or RETR.

\subsection{File Action Commands}
The final, and most effective, group of commands tested were file actions. The file actions CWD, CDUP, SMNT, and DELE were all tested on the Parrot through the ftp\_pre\_post module. The CWD command, or change working directory command, allows the user to work with a different directory or dataset. The argument for this command is a pathname specifying a directory to transfer to. The CDUP command is a special case of CWD, which allows the user to move up the filesystem tree and makes folder navigation much easier. The SMNT command, which stands for structure mount command, allows the user to mount a file system or data structure onto the current branch. The parameter this command takes is the file system or directory that is going to be mounted. SMNT, unlike CWD and CDUP, isn't permitted with anonymous FTP due to it altering the data in the file system, rather than altering how the file system is viewed. The DELE, like SMNT, isn't allowed to be executed by the user with anonymous FTP. The DELE command allows the user to delete a file or directory with the parameter being the name of a file or pathname.

\subsubsection{CMD Command}
CWD was the first of the file action commands tested on the Parrot. CWD shows minimal video distortion when part of testing. Without any parameters, CWD is able to block the Parrot's GPS connection when the size of the packets reaches 3,000 bytes. While the longevity of the test is based on how the long the test goes on for, it is relevant to mention that the effects of the test will immediately dissipate once the test is terminated. When parameters are used, however, CWD begins to act similarly to the PASV command. CWD, like the PASV command, is permitted with anonymous FTP and is therefore processed and attempted. This means that when parameters are present, the delay seen with the PASV command is seen once again. CWD will begin to cut out immediately with parameters involving a "/". To further test the delay and ignoring of the command, invalid "/"-related parameters were inputted while the test was running. With a range of packet sizes from 3,300 to 3,750 bytes, the invalid character and imaginary directories put as parameters would be processed and accepted by the UAV just as any other parameter would. This delay would happen after the test had been running for about 5 to 10 seconds and is a promising place for even more sophisticated exploits.

\subsubsection{CDUP Command}
CDUP, being so closely related to CWD behaved similarly. The GPS on the Parrot began to cut out with a packet size of 2,800 bytes. With parameters involving "/", the GPS would cut out at different ranges intermittently. From 100 to 300 bytes there are brief losses of GPS connection, from 700 to 1,100 bytes there is a continuous loss of GPS signals, and the same occurs with packets ranging in sizes from 1,200 bytes to 1,300 bytes. CDUP also showed the same delay characteristic as CWD. CDUP, after running the test for 5 to 10 seconds, will take invalid parameters and still force the Parrot to lose GPS connection. After the test is launched with CDUP or CWD, the Parrot has issues establishing a GPS connection which implies that both tests leave a long impact on the Parrot's system.

\subsubsection{SMNT and DELE Command}
SMNT and DELE were both not as effective as CWD or CDUP but still demonstrated some interesting symptoms. Unlike the CDUP and CWD command, both the SMNT and DELE command had no dramatic improvements when tested with special characters. SMNT was able to block all GPS connections when the packet size was in the range of 1,000 bytes to 5,000 bytes. SMNT also causes the Parrot to have difficulty establishing a GPS connection. The DELE command, being a simple request that requires little computation and is not permitted by anonymous FTP, showed little effect on the Parrot's behavior. There was intermediate video obscuration but the UAV's GPS connection remained consistent and strong.

\subsection{Flight Testing}
It is important to know that all the previous tests mentioned were done while the Parrot was not flying. The Parrot was online, operational, and in its landed state. The reason for this is because the Parrot treats network traffic differently when it is in flight compared to it being landed. Due to security measures implemented on the Parrot Bebop 2, any network traffic that isn't part of flying, is never executed. This means that although every FTP command will be processed, the Parrot will refuse to drain resources from subsystems critical to flight such as GPS navigation. Surprisingly, the Parrot will not try to regain GPS navigation if lost before take off. When the previously mentioned tests are launched before the Parrot has taken off, the GPS function of the UAV will remain unavailable for the entirety of the flight. The loss of GPS navigation as a feature may stay until the UAV is restarted with s utilizing effective commands like CWD, CDUP, and REIN. If the test is launched during the UAV's fight, however, the GPS connection will remain but other subsystems of the Parrot will begin to fail. Video obscuration becomes much more apparent, delay in the GPS updating the Parrot's position becomes frequent, motor speeds begin to change randomly, the charge in the Parrot's battery drains quickly, and pilot input becomes slightly delayed or entirely ignored at certain points of the flight. There was also a direct correlation seen between commands that were effective at eliminating GPS navigation and commands causing other subsystems to fail during flight.
\begin{figure}[!t]
    \includegraphics[width=\linewidth]{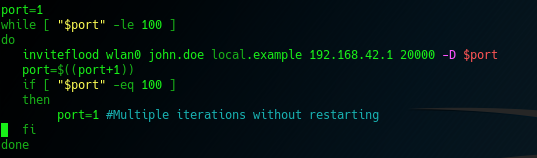}
    \caption{The bash script used to automate the InviteFlood fuzzing test on the Parrot Bebop 2}
    \label{fig:InviteFlood code}
\end{figure}

\subsection{General Fuzzing Tests}
To see if this novel and dangerous test is isolated to purely FTP anonymous, a series of tests were conducted with a tool called InviteFlood. InviteFlood is a tool that will flood session initiation protocol (SIP) packets and session description protocol (SDP) packets to any device at any port. To see if fuzzing other ports would cause GPS communication to shut down, a bash script shown in Fig. \ref{fig:InviteFlood code} was set up to call InviteFlood to fuzz a series of ports. 20,000 packets would be launched at ports 1 to 100. Low port numbers were targeted due to their popularity in data transmission. Many common network protocols use low port numbers making the range from 1 to 100 a good set of ports to test. Symptoms seen by the fuzzing tests, such as the charge in the battery dropping and motor speed irregularity was common. The loss of GPS navigation, however, was only seen in a specific range of ports. Ports 14 to 27 showed a loss of GPS connection when fuzzed, as well as ports 48 to 65. This indicates that other services running on the Parrot can be further exploited by fuzzing specific commands as well. This discovery also demonstrates how dangerous it can be to have unsecured ports on a UAV or any IoT device.

\subsection{Specifics of Fuzzing Tests}
While the termination of GPS signals on any IoT device is a serious vulnerability, the specifics of how the termination behaved is what makes this novel test so perilous. The test is executed quickly. As soon as a certain packet size was reached or a triggering special character was used, the effects of the tests were immediate. The longevity of the test is concerning in regards to the security of IoT devices. As mentioned previously, CDUP showed connectivity issues after the termination of the test. If launched before the flight of the UAV, the GPS signals would remain terminated for the entirety of flight. Such quick execution and dramatic longevity are dangerous threats to IoT device security and especially the functionality of UAVs.

\section{Identified Vulnerabilities}
The most advantageous attribute of this fuzzing test is that it exposes a plethora of new vulnerabilities. This fuzzing test not only shows the security risk of unsecured FTP anonymous but how exploited insecure ports can weaken the subsystems of UAVs. Many domestic UAVs sold at stores fall victim to even more vulnerabilities than the Parrot Bebop 2 used for testing. For that reason, it is necessary to not only diagnose how this test can be used in similarly secured UAVs but other, unsecured models of UAVs available to the public. UAVs with WPA2 encryption aren't safe either. Having a weak password can make secure UAVs susceptible to this test. The most dangerous aspect of this test is the wide variety of effects it has.

\subsection{Video Obscuration}
The most noticeable and debilitating effect brought by fuzzing tests is video obscuration. When navigating complex environments that are rapidly changing, the UAV's vision is the best way to navigate an area. The UAV's video feed is especially useful when navigating a location far away from the pilot. When the only source of knowledge regarding your exact location is camera feed, the disruption of that camera feed becomes a major issue. What's even more unfortunate is that if a fuzzing test causes the delay of the video feed, the pilot may not be able to tell if what they are seeing is still relevant. When many UAV's lose connection with a certain subsystem, they'll display what the last signal informed the UAV. While this can mean the GPS will take a while to update, it can also be as serious as the video feed freezing entirely or giving a misleading perspective. With so much information restricted from the pilot due to a fuzzing test, crashing can seem inevitable. The other crucial detail about this test is that it isn't necessarily specific to UAVs. Other IoT devices can be victims of this fuzzing test. Security cameras, automated vehicles, and vision-based sensors would all be directly affected by a fuzzing test of this magnitude.

\subsection{Battery Draining Attacks}
Video obscuration isn't the only symptom recorded of a fuzzing test on a UAV. The extremely quick drop in battery charge is a very concerning side effect as well. The most likely cause of this swift drop in power is the amount of computation needed to handle all of those requests. While managing an FTP server's constant requests may be easy on its own, the Parrot is processing a lot more at the same time. The motors, video streaming, GPS navigation, and connection with the pilot all have to be powered at the same time these requests are being flooded thought the UAV's system. The other factor that makes this effect have such a great impact, is that many of the other symptoms of fuzzing and flooding a UAV become more dramatic when the battery level is lower. That means the more the battery is drained, the more susceptible the UAV is to other attacks. Just like video obscuration, the sharp drain of battery power could be used against other IoT devices. Security systems, remote-controlled vehicles, and other IoT devices that run solely on battery power may be at risk.

\subsection{Delay in User Input and Hardware}
While uneven motor speed and a delay when controlling the Parrot are problems more specific to UAVs, it is still a notable effect of a fuzzing test. The fuzzing test being able to affect the Parrot's hardware means that there is a possibility of other IoT devices having their hardware malfunction as well. The slow reaction to user input is also a major issue with many IoT devices that are controlled and managed remotely. Both of these effects would be disastrous for remotely operated machinery. It is possible that being able to control specific malfunctions in the hardware, combined with the ability to limit user interaction with the UAV, could lead to a type of replay attack. The hacker would learn what type of data makes the UAV swerve one way or another and can eventually create a program to steer the UAV remotely while the original user is blocked off. This type of replay attack, if implemented, would be an extremely effective way to hijack a UAV or other remotely operated vehicles.

\subsection{GPS Elimination}
The most novel part of the fuzzing test is the elimination of GPS navigation. While this effect is based on the time the test is launched and what state the UAV it can lead to even more perilous attacks. A feature that many commercial UAVs possess is an autonomous flight feature. This works by generating a flight path for the UAV to follow. For such an operation to work properly, GPS needs to constantly update the UAV's position and maintain a strong connection with the UAV. \cite{HoTi2016Milcom}\cite{twente} If this test were to be launched on UAV planning to take an autonomous flight, malfunctions would incessantly plague the operation. Not only does this test prevent an autonomous flight to take place, but it also gives an attacker direct access to the UAV's GPS subsystem. It is then feasible for an attacker to send false GPS data, that the UAV would trust and follow. Giving the UAV specific locations as GPS data would be another way to launch a replay attack and eventually hijack the UAV. It is also possible that once GPS navigation fails as the UAV begins to launch, that the attacker will now have direct input into the UAV. This could allow the attacker to steal the images and video collected on the UAV, or control other aspects of the UAV. Deleting critical files from the UAV's file system, injecting malware to collect more information, or manipulating the UAV's network connections are all legitimate attacks that could be subjected to a UAV based on this fuzzing test.

Without the proper security protocols in place, many commercial UAVs will fall victim to this test. The more communication systems and processes managed by a UAV, the more likely a fuzzing test would be able to exploit this vulnerability. Many other domestic UAVs use SSH or Telnet as their primary communication subsystem and can be exploited in the same fashion. Using SSH specific commands or Telnet specific commands could have even greater implications due to limited security measures taken. Although there are many different respective models of UAVs, an unsecured way to communicate with that UAV, whether it is through FTP, Telnet, or SSH doesn't make a difference in terms of this test. The cybersecurity of commercial UAVs and IoT devices are at risk of this kind of fuzzing test without the proper security measures taken to defend them. \cite{side-channel}

\subsection{User Protection}
Unsecured UAV's and other IoT devices aren't the only possible victims of this fuzzing test. A device's security is only stable when a user follows proper protocol when maintaining its security. Users securing their data with simplistic passwords that can be easily cracked are just as vulnerable as unsecured UAVs. Many people aren't aware of how much data a hacker can obtain from an IoT device. A hacker can break into a UAV's network wouldn't just have the opportunity to attack the UAV, but the original pilot of the UAV. The fuzzing test could be used to weaken the secure connection between the UAV and the user and allow for the interception of private information that can be used for later, malicious purposes \cite{sans,drop}. UAV security isn't just about securing UAVs, it's much more than that. This fuzzing test exposes a point of weakness that could be used to infiltrate a much larger network. Exploiting the vulnerabilities of unsecured FTP, SSH, Telnet, and other communication services used by UAVs is a major security risk in the IoT world. Such a large risk is what makes this discovery such a critical topic to continue research in. 

\subsection{Diagnosing IoT Devices}
\begin{figure}[t!]
  \includegraphics[width=\linewidth]{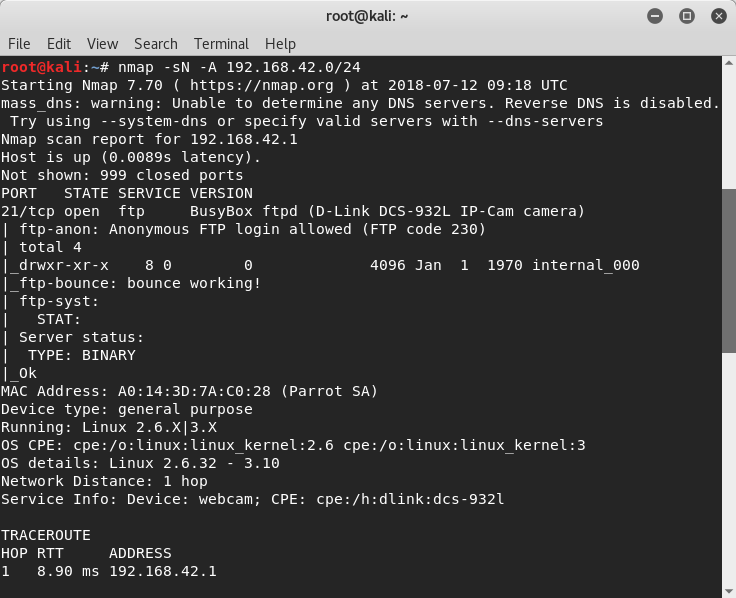}
  \caption{Nmap quickly identifying anonymous FTP service on Parrot}
  \label{fig:nmap}
\end{figure}

The confirm if this fuzzing test has the same results on other UAVs or IoT devices is very simple. A packet flooding tool like InviteFlood is being used, flooding packets to ports 1 to 100, may be the quickest way to see if a device is vulnerable to this test. For UAVs, launch the test two times. Once when it is in its landed state and once when it is in flight. If abnormalities occur, then there is a reason to investigate the services located on those ports. To look at specific services or open ports scan the device with Nmap, as seen with Fig. \ref{fig:nmap}, or manage its network traffic with a packet sniffer. Once an open service is found, look into that service's specific keywords and commands that can be sent in a fuzzing test. Launch a fuzzing test with commands that show a high possibility of overwhelming the UAV. Commands that require a large amount of computation and take a wide variety of arguments are optimal. 

\section{Proposed Countermeasures}
With IoT being a relatively young and new expansive addition to our lives, the general lack of cybersecurity with many IoT devices is extremely alarming. With security systems connected to mobile devices, personal assistants constantly listening in people's homes, and simple devices such as refrigerators and thermostats connected to the internet, it is no wonder why the IoT has become such a pertinent matter regarding cybersecurity. UAV's are no exception.  

Although the aforementioned test is effective, simple countermeasures can be implemented to either prevent or limit the extent of this test. All of these countermeasures involve changing the software of the UAV, which developers can implement quickly and efficiently. WPA2 encryption, extensive soft limits, port management, and dynamic security protocols are all ways UAVs can be defended from this test.

\subsection{WPA2 Encryption}
WPA2 encryption is the strongest line of security for all UAVs. UAVs are essentially flying access points. They generate their own network with the user and UAV at its core. Without WPA2 encryption, the secure connection between the UAV and the user is exposed to hackers. One strategy UAVs use as a security measure is to allow only phones to make a connection. While this may be a valid form of prevention for script kiddies, it is not completely secure. Tools like macchanger are capable of changing a computer's MAC address to any custom MAC address the attacker wants. Macchanger gives the attacker the ability to mask their computer as any mobile device. For that reason, WPA2 encryption is a necessary security measure to implement in commercial UAVs. The UAV that the fuzzing test was launched on, a Parrot Bebop 2, didn't have WPA2 encryption and is the reason why this vulnerability was so much easier to exploit. Of course, WPA2 encryption is only as good as the password protecting the UAV's network in the first place. Users with weak passwords that are extremely easy to crack will be just as vulnerable as a UAV with no password in the first place \cite{wpa2} \cite{networks} \cite{DePu2018VLSID}. 

\subsection{Port Security}
Just because a UAV has WPA2 encryption implemented, doesn't mean that other security measures shouldn't be taken. One of the biggest factors affecting the efficiency of this test is the management of ports on the UAV. During a flight, only critical ports should be opened for use. Services like FTP, SSH, and Telnet should all be restricted and require a second level of authentication. The only ports that should be opened for use, are those necessary for the flight of the UAV. With limited port security on UAVs, this fuzzing test can flood multiple ports at the same time and immediately distress the UAV. IoT devices are just as vulnerable. Not accounting for every access point on a device is what leads to vulnerabilities that this novel fuzzing test exploits. Forcing a device to calculate and respond to thousands of requests with no way to defend itself. Port security is a necessity for all IoT devices. Without port security, any service being run on an IoT device can be manipulated, overwhelmed, exploited, and weakened. The fact that many commercial UAVs and IoT devices don't practice port security is a major issue and is a reason for a multitude of vulnerabilities being exploited.

\subsubsection{Static vs. Dynamic Port Security}
UAV's may vary in what features are provided or how well they are secured, but port security can be fundamentally the same for all of them \cite{traffic} \cite{drop} \cite{Shashok}. A UAV may have hundreds, if not thousands, of ports that it can use. For that reason, there are two methods of maintaining port security. Static port security involves closing all extraneous ports. The UAV only uses a specific subset of ports with each port being used for exactly one function or service. Whether it is broadcasting, sharing data, or accepting input from the user, there is a port registered for each operation. Securing each of these ports, rather than securing every port on the UAV is much simpler and can avoid random overflow attacks brought by flooding. This strategy begins to show flaws when there are too many processes to handle. If a UAV is providing a suite of services, then it can be difficult to secure each port. The other problem is that the ports being used for communication will be quickly found out by packet sniffers. This information could lead to targeted attacks against specific ports. The second option is dynamic port security. Instead of having each service assigned to a unique port, have the service communicate on multiple ports. The service will switch what port to send and receive signals based on a specific algorithm. The ports that aren't being used will still be closed, but now attackers can target only one port. The attacker will have to attempt attacking multiple ports to get any results. With such a tactic used, fuzzing and flooding attacks would be much harder to launch on a large scale. If a swarm of UAVs or IoT devices used dynamic port security, then a flooding attack wouldn't be able to affect the entire swarm. With each device using a different port at what seems like random times, it would be extremely arduous for an attacker to breach the security of every, individual device.

\subsection{Soft Limits}
Port security can be taken one step further with soft limits. Soft limits are rules programmed into a UAV or any IoT device that detect whether a request sent from a node is valid. On the Parrot Bebop 2, a soft limit was used for how large a packet could be. Packets that went over this limit were rejected to protect the UAV from overflow attacks. While this method is a good start, more can be done on a software level. One vulnerability that made the fuzzing test so successful at eliminating GPS navigation was that the Parrot Bebop 2 had to understand the entire request before checking if it is permissible. It would've been beneficial if the Parrot looked at the specific keywords before computing what the rest of the command. Searching for keywords would allow the UAV to restrict commands that it knows aren't allowed. The same methodology could be applied to the arguments of commands to check if they are inherently invalid and shouldn't be fully processed.

Soft limits can be employed for other purposes besides port security as well. If UAV already knows that it is hosting one user and receives another request to join its network, soft limits would be able to automatically deny that request. This combination of user management and soft limits would allow UAVs to prevent a second user to gain access while maintaining a strong connection with the original user. Soft limits can also be used when managing IoT devices and how they manage network traffic. IoT devices can add soft limits to proficiently prevent fuzzing tests and other forms of attacks from occurring. Creative implementations of soft limits in IoT devices will prevent many overflow attacks and increase the level of cybersecurity in the IoT world.

\subsection{Adaptive Security}
With many UAVs and IoT devices, the security protocols and rules used remain constant all the time. While this isn't necessarily a bad thing, a UAV being able to recognize when an overflow attack is occurring would be an extremely useful security feature. If a UAV notices large amounts of data being sent from the same user, the UAV could notify a trusted user or pilot to give clarifying instructions to the UAV. If the trusted user indicates to the UAV that the source of these requests is unknown, the UAV can begin blocking all traffic from the address. Having adaptive security measures in place is a way to prevent many simplistic attacks that plague IoT devices. \cite{secure-drones} IoT devices could also be configured to trust only certain IP addresses. Since UAVs configure users with static IP addresses, this feature would be a simple addition to many commercial UAVs and other IoT devices.

\subsection{Necessity}
The countermeasures listed above are crucial to defending IoT devices from this fuzzing test. Adaptive security measures to identify suspicious behavior, soft limits to prevent computing pointless requests, and managing access points are all critical to UAV and IoT security. Without the proper management and implementation of security protocols, more vulnerabilities will be exploited to steal private information and gain access to other restricted devices.
\section{Conclusion}
In this paper, a novel attack involving the interruption of GPS navigation through a fuzzing test has been shown. The vulnerabilities in port security and anonymous FTP were both exploited to interrupt GPS navigation on a Parrot Bebop 2. The effects of fuzzing tests were discussed in terms of UAV security and IoT security. Possible attacks that could stem from this exploit were discussed, as well as countermeasures that commercial UAVs could implement. Commercial UAVs and other IoT devices were analyzed in great detail regarding proper network security and potential risks users may face when owning these devices. This novel attack further proves how important the network security of UAVs is and how dire of a need network security is for commercial UAVs.

%
\bibliographystyle{IEEEtran}
\bibliography{UAV-ref}

\begin{thebibliography}{10}
\providecommand{\url}[1]{#1}
\csname url@samestyle\endcsname
\providecommand{\newblock}{\relax}
\providecommand{\bibinfo}[2]{#2}
\providecommand{\BIBentrySTDinterwordspacing}{\spaceskip=0pt\relax}
\providecommand{\BIBentryALTinterwordstretchfactor}{4}
\providecommand{\BIBentryALTinterwordspacing}{\spaceskip=\fontdimen2\font plus
\BIBentryALTinterwordstretchfactor\fontdimen3\font minus
  \fontdimen4\font\relax}
\providecommand{\BIBforeignlanguage}[2]{{%
\expandafter\ifx\csname l@#1\endcsname\relax
\typeout{** WARNING: IEEEtran.bst: No hyphenation pattern has been}%
\typeout{** loaded for the language `#1'. Using the pattern for}%
\typeout{** the default language instead.}%
\else
\language=\csname l@#1\endcsname
\fi
#2}}
\providecommand{\BIBdecl}{\relax}
\BIBdecl

\bibitem{krishna}
C.~G.~L. Krishna and R.~R. Murphy, ``A review on cybersecurity vulnerabilities
  for unmanned aerial vehicles,'' in \emph{2017 IEEE International Symposium on
  Safety, Security and Rescue Robotics (SSRR)}.\hskip 1em plus 0.5em minus
  0.4em\relax IEEE SSRR Conference, Oct 2017, pp. 194--199.

\bibitem{Xingqin}
\BIBentryALTinterwordspacing
X.~Lin, R.~Wiren, S.~Euler, A.~Sadam, H.~Maattanen, S.~D. Muruganathan, S.~Gao,
  Y.~E. Wang, J.~Kauppi, Z.~Zou, and V.~Yajnanarayana, ``Mobile networks
  connected drones: Field trials, simulations, and design insights,''
  \emph{CoRR}, vol. abs/1801.10508, 2018. [Online]. Available:
  \url{http://arxiv.org/abs/1801.10508}
\BIBentrySTDinterwordspacing

\bibitem{usenix}
\BIBentryALTinterwordspacing
Y.~Son, H.~Shin, D.~Kim, Y.~Park, J.~Noh, K.~Choi, J.~Choi, and Y.~Kim,
  ``Rocking drones with intentional sound noise on gyroscopic sensors,'' in
  \emph{Proceedings of the 24th USENIX Conference on Security Symposium}, ser.
  SEC'15.\hskip 1em plus 0.5em minus 0.4em\relax Berkeley, CA, USA: USENIX
  Association, 2015, pp. 881--896. [Online]. Available:
  \url{http://dl.acm.org/citation.cfm?id=2831143.2831199}
\BIBentrySTDinterwordspacing

\bibitem{networks}
E.~Yanmaz, S.~Yahyanejad, B.~Rinner, H.~Hellwagner, and C.~Bettstetter, ``Drone
  networks: Communications, coordination, and sensing,'' \emph{Ad Hoc
  Networks}, vol.~68, 09 2017.

\bibitem{infosec}
P.~Paganini, ``Hacking drones ... overview of the main threats,''
  \url{https://resources.infosecinstitute.com/hacking-drones-overview-of-the-main-threats/},
  June 2013, accessed: 2018-7-15.

\bibitem{twente}
N.~M. Rodday, R.~d.~O.~Schmidt, and A.~Pras, ``Exploring security
  vulnerabilities of unmanned aerial vehicles,'' in \emph{NOMS 2016 - 2016
  IEEE/IFIP Network Operations and Management Symposium}, April 2016, pp.
  993--994.

\bibitem{Hartmann}
K.~Hartmann and K.~Giles, ``Uav exploitation: A new domain for cyber power,''
  in \emph{2016 8th International Conference on Cyber Conflict (CyCon)}, May
  2016, pp. 205--221.

\bibitem{sans}
D.~Kovar, ``Uav (aka drone) forensics,''
  \url{https://www.sans.org/cyber-security-summit/archives/file/summit-archive-1492184184.pdf},
  June 2016, accessed: 2018-6-25.

\bibitem{drop}
e.~Clark, Devon~R, ``Drop (drone open source parser) your drone: Forensic
  analysis of the dji phantom iii,'' in \emph{Digital Investigation},
  vol.~22.\hskip 1em plus 0.5em minus 0.4em\relax Elsevier Science Publishers,
  2017, pp. S3--S14.

\bibitem{hackBebop}
M.~L. et~al, ``An unofficial bebop hacking guide 1.6.1,''
  \url{http://fargesport
  folio.com/wp-content/uploads/2017/01/BebopHackingGuide.pdf}, 2016, accessed:
  2018-11-19.

\bibitem{traffic}
M.~A. K. W. E. T. B. D. J.~G. Wets, ``Uav-based traffic analysis: A universal
  guiding framework based on literature survey,'' in \emph{Transportation
  Research Procedia}, vol.~22, 2017, pp. 541--550.

\bibitem{ftp}
R.~Haden, ``Data network resource - ftp,'' \url{http://www.rhyshaden
  .com/ftp.htm}, 1996, accessed: 2019-1-10.

\bibitem{HoTi2016Milcom}
M.~Hooper, Y.~Tian, R.~Zhou, B.~Cao, A.~P. Lauf, L.~Watkins, W.~H. Robinson,
  and W.~Alexis, ``Securing commercial wifi-based uavs from common security
  attacks,'' in \emph{MILCOM 2016 - 2016 IEEE Military Communications
  Conference}, Nov 2016, pp. 1213--1218.

\bibitem{side-channel}
A.~K. Khan and H.~J. Mahanta, ``Side channel attacks and their mitigation
  techniques,'' in \emph{2014 First International Conference on Automation,
  Control, Energy and Systems (ACES)}.\hskip 1em plus 0.5em minus 0.4em\relax
  Curran Associates, Inc, Feb 2014, pp. 1--4.

\bibitem{wpa2}
S.~Nichols, ``Cracking wpa2 passwords just got easier,''
  \url{https://www.theregister.co.uk/2018/08/06/wpa2\_wifi\_pmkid\_hashcat/},
  August 2018, accessed: 2018-8-16.

\bibitem{DePu2018VLSID}
V.~Dey, V.~Pudi, A.~Chattopadhyay, and Y.~Elovici, ``Security vulnerabilities
  of unmanned aerial vehicles and countermeasures: An experimental study,'' in
  \emph{2018 31st International Conference on VLSI Design and 2018 17th
  International Conference on Embedded Systems (VLSID)}, Jan 2018, pp.
  398--403.

\bibitem{Shashok}
N.~Shashok, ``Analysis of vulnerabilities in modern unmanned aircraft
  systems.''\hskip 1em plus 0.5em minus 0.4em\relax Tuft University, 2017, pp.
  1--10.

\bibitem{secure-drones}
J.~H. Cheon, K.~Han, S.~Hong, H.~J. Kim, J.~Kim, S.~Kim, H.~Seo, H.~Shim, and
  Y.~Song, ``Toward a secure drone system: Flying with real-time homomorphic
  authenticated encryption,'' \emph{IEEE Access}, vol.~6, pp. 24\,325--24\,339,
  2018.

\end{thebibliography}

\end{document}